\documentclass[a4paper,12pt]{article} 
\usepackage{fullpage}
\usepackage{amsmath} 
\usepackage{amsthm} 
\usepackage{amssymb}	
\usepackage{graphicx} 
\usepackage{multicol} 
\usepackage[dvips,letterpaper,margin=0.7 in,bottom=0.5in]{geometry}
\usepackage{hyperref}
\usepackage{color}
\usepackage{setspace}
\usepackage{indentfirst} 
\usepackage{textcomp}
\usepackage{diagrams}
\usepackage{wrapfig}
\makeatletter 
\renewcommand{\maketitle} 
{ \begingroup \vskip 10pt \begin{center} \large {\bf \@title}
	\vskip 10pt \large \@author \hskip 20pt \@date \end{center}
  \vskip 10pt \endgroup \setcounter{footnote}{0} }
\makeatother 
 
\newcommand{\be}{\begin{equation}}
\newcommand{\ee}{\end{equation}}
\newcommand{\id}{\mathbf{1_d}}
\newcommand{\bkt}[1]{\left(#1\right)}
\newcommand{\sbkt}[1]{\left[#1\right]}
 
\let\baraccent=\= 
\renewcommand{\=}[1]{\stackrel{#1}{=}} 

\newtheorem{thm}{Theorem}[section]

\theoremstyle{definition}
\newtheorem{dfn}{Definition}
\theoremstyle{remark}

\numberwithin{equation}{section} 
\numberwithin{figure}{section} 
\numberwithin{table}{section} 

\newcommand{\horrule}[1]{\rule{\linewidth}{#1}} 

\title{	
\normalfont \normalsize 
\horrule{0.5pt} \\[0.4cm] 
\huge Hopf algebra and renormalization: A brief review \\ 
\horrule{2pt} \\[0.5cm] 
}

\author{Usman Naseer \\ \textit{Center for Theoretical Physics} 
\\ \textit{Massachusetts Institute of Technology}
\\ \date{\normalsize\today}
}

\begin{document}
\fontfamily{ptm}
\onehalfspacing

\maketitle

\thispagestyle{empty}

\begin{abstract}
We briefly review the Hopf algebra structure arising in the renormalization of quantum field theories. We construct the Hopf algebra explicitly for a simple toy model and show how renormalization is achieved for  this particular model.
\end{abstract}
\newpage
\tableofcontents
\section{Introduction}
The underlying Hopf algebraic strucuture of the process of renormalization was discovered first by Kreimer in \cite{Kreimer1}. Further progress was made in formulating the renormalization procedure in the language of Hopf algebra and doing explicit computations using this algebraic structure, in \cite{Kreimer2,Kreimer3,Kreimer4,Kreimer5}. The purpose of this article is to briefly review this algebraic structure. For simplicity, we avoid many of the technicalities of the quantum field theory by considering a simple toy model containing only nested divergences. Issues related to overlapping divergences and more realistic field theoretic models have been discussed in literature( see refs. \cite{Kreimer6} and \cite{QED}). For a more detailed review of this subject see \cite{review}.

This article closely follows the conventions and notation used in \cite{knots}. For a detailed treatment of renormalization procedure, see \cite{collins}. For a mathematically rigorous introduction to Hopf algebra, see \cite{qgroup}.

The article is organized as follows. In section \ref{sec:pre}, we describe necessary notation and conventions. In section \ref{sec:halgebra} we explicitly construct the Hopf algebra structure. For clarity of our arguments and construction, most of the proofs have been relegated to appendices. The article is concluded with a very simple example in appendix \ref{app:example}
\section{Preliminaries}\label{sec:pre}

\subsection{The forest formula}
A brief summary of the BPHZ renormalization procedure and the derivation of the forest formula is given in the appendix \ref{app:bphz}. The key result of BPHZ renormalization is an iterative formula (forest formula) which gives a renormalized Feynman graph in terms of the divergent graph, its subgraphs and the corresponding counter terms.  Forest formula can be written in a schematic form as follows:
\begin{eqnarray}
\Gamma_{r}&=&\overline{\Gamma} + Z_{\Gamma},\label{eq:for1}\\
\overline{\Gamma}&=&\Gamma +\sum_{\gamma \subset \Gamma} Z_{\gamma} \left(\Gamma/\gamma\right),\label{eq:for2}\\
Z_{\Gamma} &=&-t_{\Gamma}   \overline{\Gamma},\label{eq:for3}
\end{eqnarray}
where $\Gamma$ and $\Gamma_r$ are bare and renormalized graphs respectivley. $\overline{\Gamma}$ is the graph with all the subdivergences removed. The sum is over all non-empty proper forests of $\Gamma$. $Z_{\gamma}$ and $Z_{\Gamma}$  are counter terms. $t_{\Gamma}$ is a renormalization scheme dependent operator, which removes the overall divergence associated with graph $\Gamma$. To make the notion of forest precise, let $H_{1},\cdots,H_m$ be all 1PI, non overlapping divergent subgraphs of $\Gamma$, then a proper forest of $\Gamma$ is any subset of the following set:
\begin{eqnarray}
\lbrace H_1,\cdots,H_m\rbrace.
\end{eqnarray}
\subsection{Representing the graph}
\begin{figure}[h]
\begin{center}
\includegraphics[scale=.8]{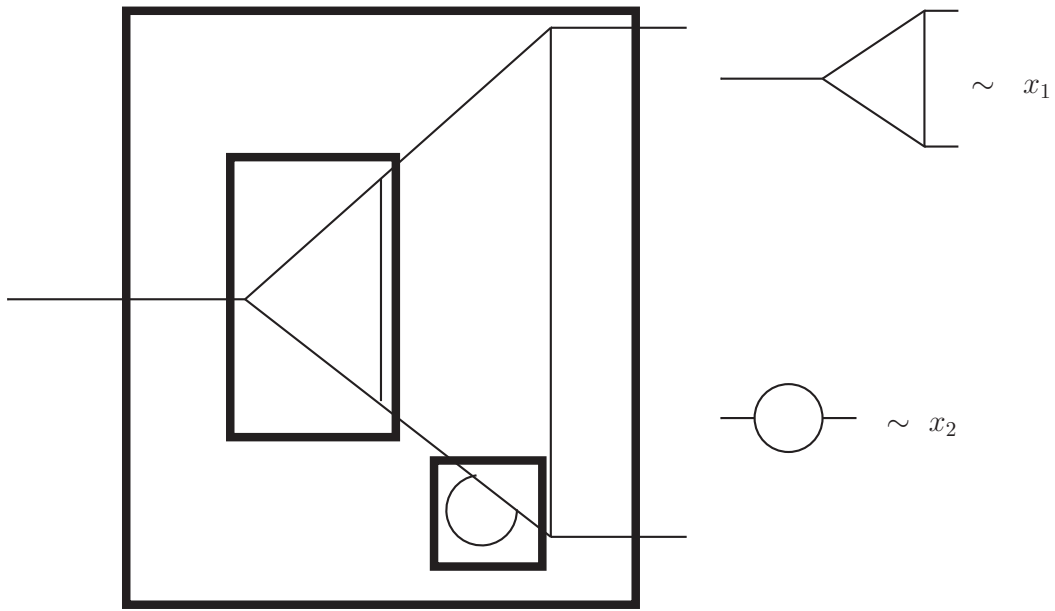}
\caption{A divergent diagram with two disjoint subdivergences}
\label{fig:TwoDiv}
\end{center}
\end{figure}

We would like to represent Feynman graphs in a more algebraic fashion such that their forest structure and subdivergences become manifest. This would be done by representing them as `parenthesized words'. Parentheses encode information about the nestedness or the disjointness of the subdivergences and letters appearing in these words correspond to graphs without subdivergences. Parenthesized words can be assigned to a graph by the following procedure:
\begin{itemize}
\item For every forest we write down a pair of brackets respecting the forest structure, i.e., if a forest $A$ is inside a forest $B$ then the pair of brackets corresponding to the forest $A$ are contained inside the pair of brackets corresponding to $B$.
\item Consider a given pair of brackets, if we shrink all the brackets/forests inside it to a point the remainder is a graph $\gamma_{i}$ without any subdivergences. We write the letter corresponding to $\gamma_{i}$ next to the right closing bracket of the pair of brackets under consideration.
\item Rest of what is contained in the pair under consideration is written to the left of this letter.
\end{itemize}
For an example, consider the diagram in figure \ref{fig:TwoDiv}. It has two disjoint subdivergences and is overall divergent when the subdivergences are shrunk to a point. The two subdivergences are contained in rectangular boxes. These subdivergences themselves are both 1PI and do not contain any subdivergences. In the figure we have also shown the letters corresponding to these subdivergences. It is easy to see that, using our rules above, this diagram corresponds to the parenthesized word  $\left(\left(x_1\right)\left(x_2\right)x_1\right)$.

Important features of this construction are following.
\begin{itemize}
\item Disjoint forests and configurations inside disjoint pair of brackets commute in this construction. i.e.,
\begin{eqnarray}
\left(\left(x_1\right)\left(x_2\right)x_1\right)=\left(\left(x_2\right)\left(x_1\right)x_1\right).
\end{eqnarray}
\item Only the forest structure of the graph is made manifest in this construction and we lose information about to which propagator or to which vertex of a graph $\gamma_j$ another graph $\gamma_i$ is attached. Several different attachment can yield the same forest structure. Hence any Feynman diagram belongs to a class given by a Parenthesized word. For example, the two diagrams in figure \ref{fig:samePW} belong to the class represented by the parenthesized word $\left(\left(x_2\right)\left(x_2\right)x_1\right)$.
\begin{figure}
\includegraphics[scale=1]{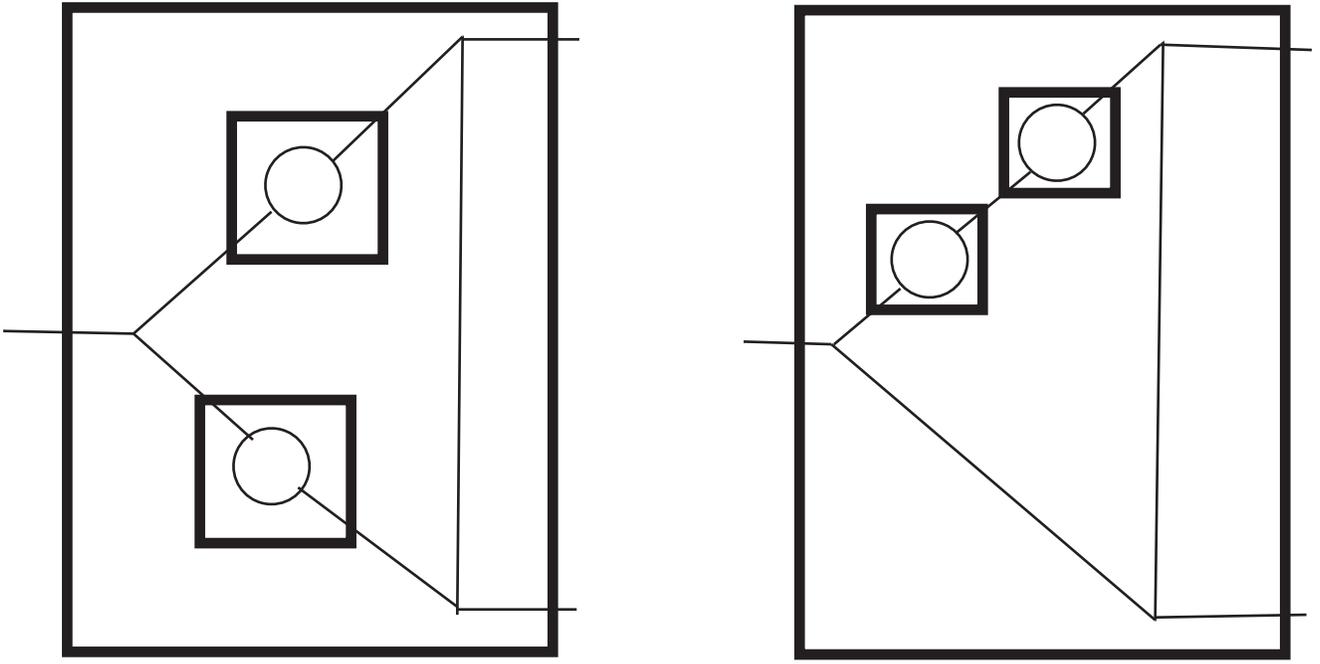}
\caption{These two Feynman diagrams corresponding to the same parenthesized word $\bkt{\bkt{x_2}\bkt{x_2}x_1}$}
\label{fig:samePW}
\end{figure}
\item A letter $x_i$ has one and only one closing bracket on its right side while it can have more than one opening brackets.
\item We include the empty graph as $\left(\right)$ which would act as a unit element (not to be confused with the unit map) in the construction of the Hopf Algebra.
\item An important characteristic of a parenthesized word is its length, which is simply the total number of letters $x_i$ appearing in it. For example, in collection (\ref{coll}), the parentheized words have lengths $0,1,2,2,3,\cdots$ respectively.

\item In general we will have a class of Feynman graphs represented by the notion of parenthesized words constructed out of letters $x_i$. Some examples are:
\begin{eqnarray}
\left(\right),\ \ \left(x_i\right),\ \ \left(\left(x_i\right)x_j\right),\ \ \left(x_i\right)\left(x_j\right),\ \ \left(\left(x_i\right)\left(x_j\right)x_k\right),\cdots\label{coll}
\end{eqnarray}
\item A parenthesized word, whose left most bracket is matched with its right most bracket is called an irreducible parenthesized word and corresponds to a 1PI Feynman graph. Examples are:
\begin{eqnarray}
\bkt{x_i},\ \ \ \bkt{\bkt{x_i}x_j},\ \ \ \bkt{\bkt{\bkt{x_i}x_j}x_k},\cdots .
\end{eqnarray}
An arbitrary irreducible parenthesized word can be represented as $\bkt{Xx_i}$, where $X$ is an any parenthesized word.
\item A parenthesized word, whose left most and the right most brackets do not match with each other is called a reducible parenthesized word and can be written as product of irreducible parenthesized words. For example
\begin{eqnarray}
\bkt{\bkt{x_i}x_j}\bkt{x_k},
\end{eqnarray}
is a reducible parenthesized word and is written as a product of two irreducible parenthesized words $\bkt{\bkt{x_i}x_j}$ and $\bkt{x_k}$.
\end{itemize}
\subsection{Hopf algebra}
\begin{wrapfigure}{r}{0.4\textwidth}
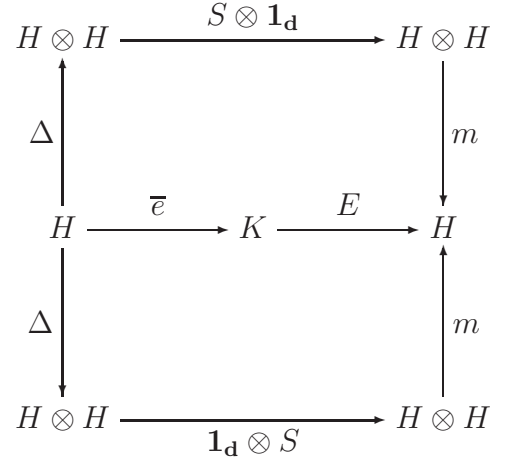

\begin{diagram}
H\otimes H &	&\rTo{S\otimes\id}	& 	&H\otimes H&\\
\uTo<\Delta &	&	&	&\dTo>m	&\\
H 	&\rTo{\overline{e}}& K	& \rTo{E} & H&\\
\dTo<\Delta	&	&	&	&\uTo>m	&\\
H\otimes H	&	&	\rTo_{\id\otimes S}&	&H\otimes H	&	\\
\end{diagram}
\caption{Commutative diagram of Hopf algebra}
\label{fig:hopf}
\end{wrapfigure}
A detailed discussion of the mathematical properties of Hopf algebra will lead us off topic. In this subsection, we will give the formal definition of a Hopf algebra and different elements appearing in the definition. We will also give a rough sketch of how the procedure of renomalization can be described by an underlying Hopf algebra structure. These notions will be made more precise in the next section.

Formally a Hopf algebra is defined as following.

\begin{dfn}\label{defHopf}
A Hopf algebra is an associative and co-associative bialbegra $H$ over a field $K$ with a K-linear map $S:\ H\rightarrow H$, called antipode such that the  diagram \ref{fig:hopf} commutes.
$E, \overline{e},m, \Delta $ are called unit, co-unit, product and co-product  maps respectively. The condition for the commutativity of the  diagram can be written algebraically as:
\begin{eqnarray}
m\left[\left(S\otimes \id\right)\Delta\left[X\right]\right]=m\left[\left(\id \otimes S\right)\Delta\left[X\right]\right]=E\circ \overline{e}\left[X\right]\label{hopfCom}
\end{eqnarray}
where $X$ is an element of Hopf algebra and $\id$ is the identity map.
\end{dfn}

Now we will give a brief overview of how renormalization would turn out to be related to the Hopf algebra structure. 
\begin{itemize}
\item Basic objects of the Hopf algebra are Feynman graphs $\Gamma$ which will be represented by the corresponding parenthesized word $X_{\Gamma}$. Representatives of the overall divergent graphs without subdivergences will be identified as the primitive elements of the Hopf algebra. All other elements $X_{\Gamma}$ can be built out of these primitive elements.
\item The co-product resolves the graph into its forests.
\begin{eqnarray}
\Delta\left[X_{\Gamma}\right]=\sum_{\text{all forests }\gamma}\ X_{\gamma}\otimes X_{\Gamma/\gamma}.
\end{eqnarray}
\item We have a renormalization map $R$, which extracts the divergent parts of a graph (depending on the renormalization scheme).
\item The antipode $S$ gives the counter term $Z_{\Gamma}$ through the renormalization map.
\begin{eqnarray}
S_{R}\left[X_{\Gamma}\right]=-R\left[X_{\Gamma}\right]-\ \sum_{\text{all non-empty proper forests }\gamma}R\left[S_R\left[X_{\gamma}\right]X_{\Gamma/\gamma}\right].
\end{eqnarray}
\item The renormalized Feynman graph will related to the term $m\left[\left(S\otimes \id\right)\Delta\left[X\right]\right]$, appearing in the condition of the commutativity. We would indeed see that $m\left[\left(S\otimes \id\right)\Delta\left[X\right]\right]=0$, expressing the fact that the we get a finite result.
\end{itemize}
\section{Construction of Hopf Algebra}\label{sec:halgebra}
In this section we will construct the Hopf algebra related to renormalization. This will be done by explicitly defining the all the maps and elements appearing in the definition (\ref{defHopf}). We will proceed in several steps, establishing algebra, co-algebra, bialgebra and finally Hopf algebra structure.

\subsection{The algebra structure}
 As discussed in the previous section, we will represent Feynman diagrams by parenthesized words. We will arrange these parenthesized words into an algebra structure here. Let $\mathcal{A}$ be the set of all parenthesized words. We regard this as a $\mathbb{Q}$ vector space. It is easy to see that $\mathcal{A}$ is a vector space over $\mathbb{Q}$. Now, we introduce a bilinear product map as follows:
 \begin{eqnarray}
&m&:\ \ \ \ \ \ \ \mathcal{A}\otimes \mathcal{A}\ \rightarrow \  \mathcal A,\\
&m& \left[X\otimes Y\right]\ \equiv \ \ XY \equiv YX,\ \ \  \forall\ X,Y \in \mathcal{A}.\label{defProd}
\end{eqnarray}
Also we have an identity element $e=\left(\right)$ which satisfies:
\begin{eqnarray}
eX=Xe=X\ \ \ \forall \ X\in \mathcal{A}.
\end{eqnarray}
 To understand the product (\ref{defProd}) consider the example with $X=\left(\left(x\right)x\right)$ and $Y=\left(y\right)$ then $XY$ is a well defined product given by $\left(\left(x\right)x\right)\bkt{y}$, i.e., the product of two parenthesized words give a reducible parenthesized word. By introducing the product we have furnished $\mathcal{A}$ with an algebra structure.
 
 Now we define a homomorphism (the unit map) from $\mathbb{Q}$ to the set $\mathcal{A}$ as follows:
 \begin{eqnarray}
&E&:\ \ \ \ \ \ \ \mathbb{Q}\ \to \mathcal{A},\\
&E\left[q\right]&\ \equiv e,\ \ \  \forall\ \text{rational numbers }q.\label{defunit}
\end{eqnarray}
Now, by definition, the bilinear product $m$ is associative, our algebra $\mathcal{A}$ has an identitiy element $e$ and we have constructed a homomorphism from the field of rational numbers $\mathbb{Q}$ to algebra $\mathcal{A}$, this means that the set $\mathcal{A}$ is a unital associative algebra.

\subsection{The coalgebra structure}
\begin{figure}[h]
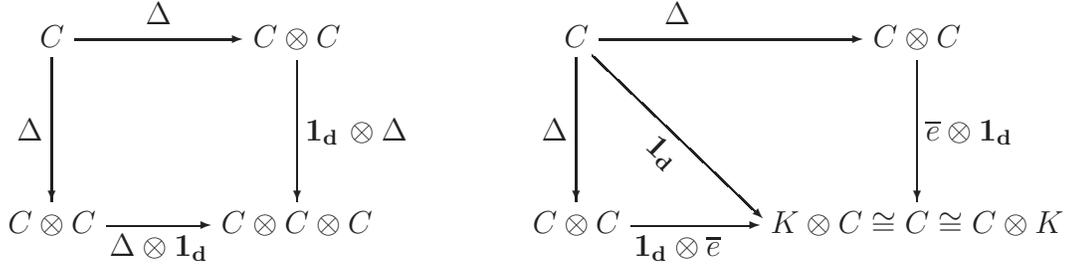

\begin{diagram}
C	&\rTo^{\Delta}&C\otimes C 	&	&	&C&\rTo^{\Delta}&	&C\otimes C\\
\dTo<\Delta &	&	\dTo>\id\otimes\Delta	&	&	&\dTo<\Delta&\rdTo_\id&	&\dTo> {\overline{e}\otimes \id}	&\\
C\otimes C&\rTo_{\Delta\otimes \id}&C\otimes C\otimes C &	&	&C\otimes C &\rTo_{\id\otimes\overline{e}}&	&K\otimes C\cong C\cong C\otimes K
\end{diagram}
\caption{Commutative diagrams for coalgebra $C$}
\label{fig:coalg}
\end{figure}

In this subsection, we furnish $\mathcal{A}$ with the structure of a coalgebra. Let us first give the formal definition of a coalgebra.
\begin{dfn}\label{defCoalg}
A coalgebra, $C$ over a field $K$ is a vector space $C$ over $K$ together with linear maps $\overline{e}:\ C\to K$ (counit) and $\Delta:\ C\to C\otimes C$ (coproduct) such that
\begin{eqnarray}
\bkt{\id\otimes\overline{e}}\Delta&=&\bkt{\overline{e}\otimes\id}\Delta,\label{coalg1}\\
\bkt{\Delta\otimes\id}\Delta&=&\bkt{\id\otimes\Delta}\Delta.\label{coalg2}
\end{eqnarray}
where $\id$ is the identity map on $C$, or quivalently, the two diagrams in figure \ref{fig:coalg} commute. In the second diagram, we have identified the naturally isomorphic spaces $C$, $C\otimes K$,$K\otimes C$.
The second equation above is also called the coassociativity condition for the coproduct $\Delta$.
\end{dfn}

Now, we will define the counit and the coproduct maps for the set $\mathcal{A}$ under consideration.

\subsubsection*{The counit}
We define a counit by:
\begin{eqnarray}
&\overline{e}&:\ \ \ \ \ \mathcal{A}\to \mathbb{Q},\\
&\overline{e}\left[e\right]&\equiv \ 1,\\
&\overline{e}\left[X\right]&\equiv \ 0,\ \ \forall\ X\neq e, \in \mathcal{A}.\ \ \ \ \ \ \ \ \ \ \ \label{defcounit}
\end{eqnarray}
This definition is motivated by the fact that there is no rational number which should be assigned naturally to an arbitrary parenthesized word and thus the counit annihilates Feynman graphs. On the other hand we assign the rational number $1$ to the empty graph $e$.
\subsubsection*{The coproduct}
The definition of the coproduct is more involved as compared to the elements defined so far. Roughly speaking, coproduct yields a sum of terms $\sum_{i}X_i\otimes Y_i$, where the first terms, $X_i$, are to be identified with divergent subgraphs and the second terms, $Y_{i}$, correspond to the remainder of the graph obtained by reducing $X_{i}$ to a point.

To give a rigorous definition of the coproduct,  it will be useful to define a projection map $P$ as follows:
\begin{eqnarray}
&P&:\ \ \ \ \ \mathcal{A}\otimes\mathcal{A}\ \to \ \mathcal{A}\otimes \mathcal{A},\\
&P&\ \equiv \left(\id-E\circ\overline{e}\right)\otimes \id.
\end{eqnarray}
It is easy to confirm the following properties of the map $P$ by explicit computation.
\begin{eqnarray}
P\left[e\otimes X\right]&=&0,\ \ \forall\ X\in \mathcal{A},\\
P\left[X\otimes Y\right]&=&X\otimes Y,\  \ \ \ \forall\  X\neq e,\ Y, \in \mathcal{A},\\
P^2&=&P.
\end{eqnarray}

We also define a useful endomorphism $B_{\left(x_{i}\right)}$, which is parametrized by a single letter $x_i$, corresponding to a primitive graph.
\begin{eqnarray}
&B_{\left(x_i\right)}&:\ \ \ \ \ \mathcal{A}\to \mathcal{A},\\ 
&B_{\left(x_i\right)}\left[X\right]&\equiv \left(Xx_i\right).
\end{eqnarray}
For example, $B_{\left(x_1\right)}\left[\left(x_2\right)\right]=\left(\left(x_2\right)x_1\right)$. With the help of the maps $P$ and $B$, we are now in a position to define the coproduct as follows.
\begin{eqnarray}
&\Delta&:\ \ \ \ \ \ \mathcal{A}\to \mathcal{A}\otimes \mathcal{A} ,\\
&\Delta\left[e\right]&\ \equiv\ e\otimes e,\\
&\Delta\left[\left(Xx_i\right)\right]&\ \equiv\ \left(Xx_i\right)\otimes e+e\otimes\left(Xx_i\right)+\left(\id\otimes B_{\left(x_i\right)}\right)\left[P\left[\Delta\left[X\right]\right]\right].\label{coprod}
\end{eqnarray}
This definition of the coproduct is complete. It is easy to use the above definition to show an  important property of the coproduct.
\begin{eqnarray}
&\Delta\left[\left(x_i\right)\right]&\ \equiv\ \left(x_i\right)\otimes e + e\otimes\left(x_i\right).\label{coprodPrim}
\end{eqnarray}
Another important property of the coproduct is:
\begin{eqnarray}
&\Delta\left[XY\right]&\  \equiv \ \Delta\left[X\right]\Delta\left[Y\right].\label{compatcoprod}
\end{eqnarray}
This can also be shown by using the definition (\ref{coprod}), however the proof is a bit involved. The proof is based on the standard induction argument on the length of the words $X$ and $Y$. 

Another way to write the coproduct is by using the Sweedler's notation, $\Delta\left[X\right]=\sum_X X_1\otimes X_2$, where the sum is over the subwords $X_1$ of $X$ and $X_2=X/X_1$. Proof of this assertion is given in appendix \ref{app:Sweed}. Using this notation and the properties of the map $P$, we can write the equation (\ref{coprod}) of the coproduct as:
\begin{eqnarray}
\Delta\left[\left(Xx\right)\right]=\left(Xx\right)\otimes e +\left(\id\otimes B_{\left(x\right)}\right)\left[\sum_X X_1\otimes X_2\right]\label{coprodSweed}.
\end{eqnarray}

Let us now consider a few example to explain how the coproduct acts on the elements of the set $\mathcal{A}$.
\begin{enumerate}
\item \begin{eqnarray}
\Delta\sbkt{\bkt{\bkt{x_i}x_j}}&=&\bkt{\bkt{x_i}x_j}\otimes e + e\otimes \bkt{\bkt{x_i}x_j}+\bkt{\id\otimes B_{\bkt{x_j}}}P\bkt{\Delta\sbkt{\bkt{x_i}}},\\
&=&\bkt{\bkt{x_i}x_j}\otimes e + e\otimes \bkt{\bkt{x_i}x_j}+\bkt{\id\otimes B_{\bkt{x_j}}}\sbkt{\bkt{x_i}\otimes e},\\
&=&\bkt{\bkt{x_i}x_j}\otimes e + e\otimes \bkt{\bkt{x_i}x_j}+\bkt{x_i}\otimes \bkt{x_j}.\label{eq:exampleDel1}
\end{eqnarray}
\item
Using the similar method (but after more tedious algebra) we can also compute:
\begin{eqnarray}
\Delta\sbkt{\bkt{\bkt{x_i}\bkt{x_j}x_k}}=&&\bkt{\bkt{x_i}\bkt{x_j}x_k}\otimes e + e\otimes \bkt{\bkt{x_i}\bkt{x_j}x_k}+\bkt{x_i}\otimes \bkt{\bkt{x_j}x_k}\nonumber \\ &&+\bkt{x_j}\otimes \bkt{\bkt{x_i}x_k}+\bkt{x_i}\bkt{x_j}\otimes \bkt{x_k}.\label{eq:exampleDel2}
\end{eqnarray}
\end{enumerate}

\subsubsection*{Coalgebra check}
We have defined the counit and the coproduct maps for $\mathcal{A}$, but in order to furnish the coalgebra structure on $\mathcal{A}$ we need to show that these maps satisfy the equations (\ref{coalg1}) and (\ref{coalg2}). The first of these relations is trivial to show due to the definition of the counit as :
\begin{eqnarray}
\bkt{\id\otimes\overline{e}}\Delta\sbkt{X}=X=\bkt{\overline{e}\otimes\id}\Delta\sbkt{X}.
\end{eqnarray}

Next, we want to show the equation (\ref{coalg2}) holds. This can be proved using induction on the length of the words. A detailed proof is given in the appendix \ref{app:coalg}.

After successfully defining a counit and a coproduct on $\mathcal{A}$, we have completed the construction of the coalgebra structure on $\mathcal{A}$. We have already established the fact that $\mathcal{A}$ is a unital coassociative algebra. The property (\ref{compatcoprod}) ensures that the algebra and the coalgebra structures are compatible. This implies that $\mathcal{A}$ is actually a bialgebra.
\subsection{The antipode}
To complete the construction of the Hopf algebra, what remains to find is an antipode. It turns out that antipode is actually the object which achieves the renormalization, it combines the terms generated by the coproduct and combines them in a way which is similar to the forest formula. We define the antipode as follows:
\begin{eqnarray}
S:&&\ \ \ \ \ \ \mathcal{A}\to \mathcal{A},\\
S\sbkt{e}&=&e,\\
S\sbkt{\bkt{x_i}}&=&-\bkt{x_i},\\
S\sbkt{XY}&=&S\sbkt{Y}S\sbkt{X},\\
S\sbkt{\bkt{Xx_i}}&=&-\bkt{Xx_i}-m\sbkt{\bkt{S\otimes\id}P_2\bkt{\Delta\sbkt{\bkt{Xx_i}}}},\label{eq:Sdef1}\\
S\sbkt{\bkt{Xx_i}}&=&-\bkt{Xx_i}-m\sbkt{\bkt{\id\otimes S}P_2\bkt{\Delta\sbkt{\bkt{Xx_i}}}},\label{eq:Sdef2}\\
P_2&\equiv& \bkt{\id-E\circ\overline{e}}\otimes \bkt{\id-E\circ\overline{e}}\equiv P_1\otimes P_1.
\end{eqnarray}
This completely defines the antipode. However, we need to show that this antipode is actually well defined and induces a Hopf algebra structure. This amounts to showing that equations (\ref{eq:Sdef1}) and (\ref{eq:Sdef2}) are equivalent\footnote{The two definitions correspond to recursive and non-recursive form of the forest formula} and also the condition (\ref{hopfCom}) is satisfied. Equivalence of the two definitions follow from the associativity of the product $m$ and the coassociativity of the coproduct $\Delta$. The detailed proof is given in appendix \ref{app:antipodeDef}. The proof that the condition (\ref{hopfCom}) is satisfied, is given in appendix \ref{app:hopfcheck}.
We have now completely furnished the set of all Feynman diagrams, $\mathcal{A}$ with the structure of a Hopf algebra. We have not yet discussed precisely how the renomalization is achieved by this structure. This will be the subject of the next section.
\subsection{From Hopf algebra to the forest formula}
In this section, we describe how the Hopf algebra constructed above produces the forest formula, generates counter terms and the renormalized Feynman graphs. We will see that an important ingredient in this regard is the renormalization map, $R$, which is renormalization scheme dependent.

Given a Feynman graph $\Gamma$, we associate a parenthesized word $X_{\Gamma}$ to it. Using the Feynman rules we obtain an integral expression associated with the graph $\Gamma$, denote it by $\phi\bkt{X_{\Gamma}}$ $\in V$, where $V$ is a vector space, endowed with suitable structure which is not important for our considerations. For example, it could be the space of Laurent polynomials in the regularization parameter. These Feynman integrals are subject to some renormalization conditions which are described by renormalization map $R:\ \ V\to V$. The renormalization map depends on the renormalization scheme, for example, in the case of minimal subtraction, $R$ picks out the only the divergent part of $\phi\bkt{X_{\Gamma}}$. The map $\phi$, the renormalization map $R$ and the antipode of the Hopf algebra $S$ give rise to a map $S_{R}$ at the level of the Feynman integrals, which is written as:
\begin{eqnarray}
S_{R}\sbkt{\bkt{Xx}}=-R\sbkt{\phi\bkt{\bkt{Xx}}}-R\sbkt{m\sbkt{\bkt{S_{R}\otimes\phi}P_2\bkt{\Delta\sbkt{\bkt{Xx}}}}}.\label{eq:antipode2}
\end{eqnarray}
with $S_{R}\sbkt{e}=e$. This map $S_{R}$ gives the counter terms for a given graph depending on the particular renormalization scheme $R$.  Consider the following examples where, for simplicity, we omit writing $\phi$ explicitly:
\begin{enumerate}
\item 
\begin{eqnarray}
S_{R}\sbkt{\bkt{\bkt{x_i}x_j}}=-R\sbkt{\bkt{\bkt{x_i}x_j}}-R\sbkt{m\sbkt{\bkt{S_R\otimes \id}P_2\bkt{\Delta\sbkt{\bkt{\bkt{x_i}x_j}}}}}.
\end{eqnarray}
We can use $\Delta\sbkt{\bkt{\bkt{x_i}x_j}}$ as computed in equation (\ref{eq:exampleDel1}). Since, $P_1$ annihilates $e$, we finally find that:
\begin{eqnarray}
S_{R}\sbkt{\bkt{\bkt{x_i}x_j}}=-R\sbkt{\bkt{\bkt{x_i}x_j}}+R\sbkt{R\sbkt{\bkt{x_i}}\bkt{x_j}}.
\end{eqnarray}
\item Similarly, after a straightforward but tedious computation one can find that:
\begin{eqnarray}
S_{R}\sbkt{\bkt{\bkt{x_i}\bkt{x_j}x_k}}&=&-R\sbkt{\bkt{\bkt{x_i}\bkt{x_j}x_k}}+R\sbkt{R\sbkt{\bkt{x_i}}\bkt{\bkt{x_j}x_k}}+R\sbkt{R\sbkt{\bkt{x_j}}\bkt{\bkt{x_i}x_k}}\nonumber \\&&-R\sbkt{R\sbkt{\bkt{x_i}}R\sbkt{\bkt{x_j}}\bkt{x_k}}.\label{eq:example}
\end{eqnarray}
\end{enumerate}
Let us now proceed further to show that the forest structure in equations (\ref{eq:for1},\ref{eq:for2},\ref{eq:for3}) emerges from the Hopf the algebra structure. 

Let $U$ be a subword of $X$, then by using the representation of the coproduct in Sweedler's notation and the fact that $P_1$ annihilates $e$, the antipode can be written as:
\begin{eqnarray}
S\sbkt{X}=- X - \sum_{U\neq e, X} S\sbkt{U}\bkt{X/U}.
\end{eqnarray}
If the parenthesized word $X$ is associated to a Feynman graph $\Gamma$ then the subwords $U\neq e, X$ are associated to the proper forest $\gamma$ of the graph $\Gamma$. Using this fact, we can now write the map $S_{R}$ in the following way:
\begin{eqnarray}
S_{R}\sbkt{\Gamma}&=&-R\sbkt{\Gamma}-\sum_{\text{proper forest }\gamma\subset \Gamma} R\sbkt{S_{R}\sbkt{\gamma}\Gamma/\gamma},\\
&=&-R\sbkt{\Gamma+\sum_{\text{proper forest }\gamma\subset \Gamma} S_{R}\sbkt{\gamma}\Gamma/\gamma}.
\end{eqnarray}
Now, if we identify $S_{R}\sbkt{\gamma}$ with counter term associated to the subgraph $\gamma$, then the argument of the map $R$ in the above equation is just $\overline{\Gamma}$, the graph $\Gamma$ with all its subdivergences renormalized as defined in equation (\ref{eq:for2}). We can also identify the renormalization map $R$ with the operator $t_{\Gamma}$, both are renormalizatio scheme dependent operators and picks out just the divergent part of a Feynman integral in MS scheme. With this identification, we see that $S_{R}\sbkt{\Gamma}$ just gives the counter term $Z_{\Gamma}$ and we recover the forest structure of equations (\ref{eq:for1},\ref{eq:for2},\ref{eq:for3}). The renormalized Feynman graph $\Gamma_{ren}$ is obtained as follows. Let $X$ be the parenthesized word associated with the graph  $\Gamma$ (We will use the parenthesized word $X$, the corresponding graph $\Gamma$ and the corresponding Feynman integral $\phi\bkt{\Gamma}$ interchangeably) then:
\begin{eqnarray}
m\sbkt{S_R\otimes \phi }\Delta\sbkt{X}&=&m\sbkt{S_{R}\otimes \phi}\bkt{e\otimes X + X\otimes e+ \sum_{\text{subwords } U\neq e,X}U\otimes \bkt{X/U}},\\
&=&\phi\sbkt{X}+S_R\sbkt{X}+\sum_{\text{subwords }U\neq e, X } S_{R}\sbkt{U}\phi\sbkt{X/U},
\end{eqnarray}
in the last equation, the first term is just the Feynman integral associated with graph $\Gamma$, the second term is the counter term $Z_{\Gamma}$ and the last term just removes the subdivergences as we have seen earlier. Now, we omit writing $\phi$ and replace parenthesized words with the respecting graphs to find:
\begin{eqnarray}
m\sbkt{S_R\otimes \phi }\Delta\sbkt{X}=\Gamma +Z_{\Gamma}+\sum_{\text{proper forests }\gamma\subset \Gamma}Z_{\gamma}\Gamma/\gamma=\overline{\Gamma}+Z_{\Gamma}=\Gamma_{ren}.
\end{eqnarray}
Earlier, we showed that at the Hopf algebra level, the operator $m\sbkt{\bkt{S\otimes \id}\Delta}$ annihilates any parenthesized word other than the unit $e$. This expresses the fact that at the level of the Feynman integrals we will get essentially a finite result.

\section{Summary}\label{sec:sum}
In this section we will briefly summarize the key results of this article.
By representing the Feynman diagrams as parenthesized words, we furnished them into a set $\mathcal{A}$. We also included the empty graph, represented by the unit element $e$, in that set. Then we introduced an algebra structure on $\mathcal{A}$ by defining a bilinear product $m:\mathcal{A}\to \mathcal{A}$. We also defined a unit map $E:\mathbb{Q}\to \mathcal{A}$, furnishing $\mathcal{A}$ into a unital associative algebra. Next, we introduced the coalgebra structure on $\mathcal{A}$ by defining the counit map and the coproduct map. The coproduct was defined in such a way that it was compatible with the product $m$ and hence we obtained a bialgebra structure on $\mathcal{A}$. To complete the construction of the Hopf algebra, we defined an antipode map $S:\mathcal{A}\to\mathcal{A}$. We also showed that the struture of the forest formula is recovered if we identify the antipode with the counter term of a specific graph. To make this notion precise, we defined a map $\phi:\mathcal{A}\to V$, which assigns a parenthesized word an analytic expression (Feynman integral) using the Feynman rules. We defined the renormalization map $R$ which gives the divergent part of a Feynman integral. It turned out antipode $S$ induced the counter term for a graph via $R$.

The most important result we obtained is the equivalence of the antipode and the forest formula. This equivalence followed by making a set of identifications between the elements of the Hopf algebra and the objects of the standard renormalization theory. We list these identifications here.
\begin{itemize}
\item 1PI Feynman graph $\Gamma$ with subdivergences are identified with irreducible parenthesized word $(Xx)$ whose bracket structure matches the forest structure of $\Gamma$, and the letters label the components of $\Gamma$ obtained after reducing the subdivergences to a point.
\item The counter term $Z_{\Gamma}$ is identified with $S_{R}\sbkt{\Gamma}$.
\item The Feynman graph, with all its subdivergences renormalized, $\overline{\Gamma}$ is identified with the object:
\begin{eqnarray}
m\sbkt{\bkt{S_R\otimes \phi}P_{R}\Delta\sbkt{\bkt{Xx}}},\ \ \text{where}\ \ \ P_R=\id\otimes\bkt{\id-E\circ\overline{e}}.
\end{eqnarray}
\item The renormalized Feynman graph $\Gamma_{ren}=\overline{\Gamma}+Z_{\Gamma}$ is identified with:
\
\begin{eqnarray}
m\sbkt{\bkt{S_{R}\otimes\phi}\Delta\sbkt{\bkt{Xx}}}.
\end{eqnarray}
\end{itemize}
\appendix
\section{BPHZ Renormalization}\label{app:bphz}

Consider a Feynman graph $\Gamma$. By using Feynman rules we can obtain the corresponding analytic expression $F_{\Gamma}$. In general this expression can be written as a Laurent series in the regularization parameter $\epsilon$. If we consider $\phi$-cubed theory in $6$ spacetime dimensions and use dimensional regularization then
\begin{eqnarray}
F_{\Gamma}\ \equiv\ \sum_{n=-N}^{\infty} a_{n}\epsilon^{n} \label{Fgamma},
\end{eqnarray}
where $a_{n}$ are some coefficients and the integer $N$ is bounded above by the number of loops in the graph $\Gamma$, which can be shown explicitly. We stress here that in the general argument for the BPHZ renormalization nothing depends crucially on the particular toy model chosen here. Let us now define a `subtraction' operator associated with the graph $\Gamma$ as follows
\begin{eqnarray}
t_{\Gamma} F_{\Gamma}\ \equiv\ \sum_{n=-N}^{-1} a_{n}\epsilon^{n} \label{SubOp},
\end{eqnarray}
i.e., it picks out the divergent part of $F_{\Gamma}$. In general, the subtraction operator is renormalization scheme dependent, here we have chosen the minimal subtraction scheme. The finite part of the graph can now be written as:
\begin{eqnarray}
F_{\Gamma}^{r}\ =\ \left(1-t_{\Gamma}\right) F_{\Gamma}.
\end{eqnarray}
So, we see that the term `$-t_{\Gamma}F_{\Gamma}$' provides the counter term for the graph $\Gamma$ and $1-t_{\Gamma}$ removes the divergence associated with graph $\Gamma$ and makes it finite in the $\epsilon\rightarrow 0$ limit.

Now, consider the graph $\Gamma$ to have proper 1PI subgraphs $H_i,\ i=1,\cdots, m.$ For simplicity, we assume that all these subgraphs are overall divergent, if they are not divergent, there is no need for renormalization. We order these graphs such that if $H_{i}\subset H_{j}$ then $i<j$. Now we define the following:
\begin{eqnarray}
\overline{R}_{\Gamma} F_{\Gamma}\ \equiv \left(1-t_{H_m}\right)\cdots \left(1-t_{H_1}\right)=\left(\prod_{H_i\subset G} \left(1-t_{H_i}\right)\right) F_{\Gamma},\label{SubDivR}
\end{eqnarray}
where the product in the second equality needs to be ordered. Since the operator `$1-t_{H_i}$' removes the divergence associated with the subgraph $H_i$, we see that equation (\ref{SubDivR}) is nothing but the graph $\Gamma$ with all its subdivergences renormalized. Now we define the `Bogoliubov $R$ operator' which removes the over all divergence associated with $\Gamma$ and renders it finite:
\begin{eqnarray}
R_{\Gamma} &\equiv& \left(1-t_{\Gamma}\right)\overline{R}_{\Gamma},\\
\Rightarrow R_{\Gamma} F_{\Gamma}&=&\left(1-t_{\Gamma}\right)\ \left(\prod_{H_i\subset G } \left(1-t_{H_i}\right)\right) F_{\Gamma}. \label{RenGam}
\end{eqnarray}
Let us now define a restricted graph $\Gamma /H$ as the graph obtained by reducing $H$ to a point inside $\Gamma$, then it is easy to see that $-t_H\  F_{\Gamma}=\left(-t_HF_H\right)F_{\Gamma/H}$, i.e., we can replace the subgraph $H$ in $\Gamma$ by $\Gamma/H$ and multiply by the counter term which makes $H$ finite. We can write equation (\ref{RenGam}) as:
\begin{eqnarray}
R_{\Gamma} F_{\Gamma}&=&\left(1-t_{\Gamma}\right)\ \left( F_{\Gamma}+\sum_{\phi}\left(\prod_{H\in \phi } \left(-t_{H}\right)\right) F_{\Gamma}\right),\label{RenF2}
\end{eqnarray}
where the sum is taken over all subgraphs of $\Gamma$ (i.e., all non empty subsets (denoted by $\phi$) of the set $\lbrace H_1,\cdots, H_m\rbrace$).  
We will also need the following theorem due to Hepp \cite{hepp1966}, which we state here without proof.
\begin{thm}\label{hepp}
Let $H_1,\cdots,H_j$ be overlapping 1PI subgraphs of $\Gamma$. Then consider a subgraph $H_{12\cdots j}$ such that $H_{i}\subset H_{12\cdots j}, \forall i=1,\cdots,j$ then \begin{eqnarray}
\left(1-t_{H_{12\cdots j}}\right) t_{H_1}\cdots t_{H_j}\ =\ 0,
\end{eqnarray}
i.e., the finite part of the graph left after replacing the overlapping subdivergences is zero.
\end{thm}
Courtesy this theorem we can restrict the $\phi$ in equation (\ref{RenF2}) to be the subset of non overlapping 1PI divergences. Since $\left(\prod_{H\in \phi } \left(-t_{H}\right)\right) F_{\Gamma}$ provides the counter term associated with the subgraph $\phi$ we can write:
\begin{eqnarray}
R_{\Gamma} F_{\Gamma}&=&\left(1-t_{\Gamma}\right)\ \left( F_{\Gamma}+\sum_{\phi} Z_{\phi} F_{\Gamma/\phi}\right),\label{RenF3}
\end{eqnarray}
where $Z_{\phi}$ is the counter term which makes the subgraph $\phi$ finite. The subgraph $\phi$ is formally defined as:
\begin{eqnarray}
\phi=\lbrace H_{i}|H_{i}\subset G, H_{i} \text{ are non overlaping, 1PI} \rbrace,
\end{eqnarray} and is called a `forest' of graph $\Gamma$. In the above expression the term inside second set of parenthesis is the graph $\Gamma$ with all non-overlapping subdivergences renormalized. The remaining divergence is then removed by the operator $\left(1-t_{\Gamma}\right)$. Equation (\ref{RenF3}) is called `Zimmermann's Forest Formula'. We can write the forest formula in a schematic fashion as follows:
\begin{eqnarray}
\Gamma_{r}&=&\overline{\Gamma} + Z_{\Gamma},\label{eq:for1A}\\
\overline{\Gamma}&=&\Gamma +\sum_{\gamma \subset \Gamma} Z_{\gamma} \left(\Gamma/\gamma\right),\label{eq:for2A}\\
Z_{\Gamma} &=&-t_{\Gamma}   \overline{\Gamma},\label{eq:for3A}
\end{eqnarray}
where $\Gamma$ and $\Gamma_r$ are bare and renormalized graphs respectivley. $\overline{\Gamma}$ is the graph with all the subdivergences removed. $\gamma$ denotes all proper forests of $\Gamma$. $Z_{\gamma}$ and $Z_{\Gamma}$  are the counter terms.
\subsubsection*{Example}
We now consider an example which explains some important aspects of the forest structure of a Feynman graph and the application of the forest formula. Let us look at the diagram in figure \ref{fig:TwoDiv}. This graph (say $\Gamma_{1}$) has only two non overlapping 1PI subgraphs,  say $H_1$ and $H_2$, as labeled and boxed in the diagram. The corresponding proper forests are:
\begin{eqnarray}
\gamma_{1}=\lbrace H_1\rbrace,\ \ \gamma_{2}=\lbrace H_2\rbrace,\ \ \gamma_{3}=\lbrace H_1,H_2 \rbrace.
\end{eqnarray}
So we find that:
\begin{eqnarray}
\overline{\Gamma_1}&=&\Gamma_1+Z_{\gamma_1} \Gamma_1/\gamma_1 + Z_{\gamma_2}\Gamma_1/\gamma_2+Z_{\gamma_3}\Gamma_1/\gamma_3,\\
Z_{\gamma_1}&=&-t_{\gamma_1}\gamma_1,\ \ Z_{\gamma_2}=-t_{\gamma_2}\gamma_2,\ \ Z_{\gamma_3}=Z_{\gamma_1}Z_{\gamma_2},\\
Z_{\Gamma_1}&=&-t_{\Gamma_1}\overline{\Gamma_1},\\
\Gamma_{1r}&=&\overline{\Gamma_1}+Z_{\Gamma_1}.
\end{eqnarray}
We shoowed earlier that this diagram corresponds a parenthesized word $\bkt{\bkt{x_1}\bkt{x_2}x_1}$. If we compare the structure of the counter term $Z_{\Gamma}$ obtained here with equation (\ref{eq:example}) (which computes $S_{R}\sbkt{\bkt{\bkt{x_1}\bkt{x_2}x_1}}$), we see that the two objects have exactly the same structure after the identifications described in the section \ref{sec:sum}. 
\section{Proofs}
\subsection {Sweedler's Notation}\label{app:Sweed}
 Let $U$ be any subword of a parenthesized word $X$, then our coproduct is defined in such a way that:
\begin{eqnarray}
\Delta[X]=\sum_{ U }U\otimes \bkt{X/U}.
\end{eqnarray}
This assertion is easy to prove using the induction on length of the words. It is obviously true for words of length $1$.  Assume that it is true for word $X$ of length $n$ and then induce. Let us consider an irreducible parenthesized word $\bkt{Xx}$ of length $n+1$.
\begin{eqnarray}
\Delta\sbkt{\bkt{Xx}}&=&\bkt{Xx}\otimes e +e\otimes \bkt{Xx}+\bkt{\id\otimes B_{\bkt{x}}}P\Delta\sbkt{X},\\
&=&\bkt{Xx}\otimes e +e\otimes \bkt{Xx}+\bkt{\id\otimes B_{\bkt{x}}}\bkt{\sum_{\text{all subwords } U\neq e \text{ of X}}U\otimes \bkt{X/U}},\\
&=&\bkt{Xx}\otimes e +e\otimes \bkt{Xx}+\bkt{\sum_{\text{all subwords } U\neq e \text{ of X}}U\otimes \bkt{X/U x}} ,\\
&=&\bkt{Xx}\otimes e +\bkt{\sum_{\text{all subwords } U \text{ of X}}U\otimes \bkt{X/U x}} ,\\
&=&\sum_{\text{all subwords } U \text{ of }\bkt{Xx}}U\otimes \bkt{Xx}/U 
\end{eqnarray}
which proves our assertion for irreducible word of length $n+1$. For an arbitrary word $XY$ of length $n+1$, the assertion follows by using the induction assumption and the fact that $\Delta\sbkt{XY}=\Delta\sbkt{X}\Delta\sbkt{Y}$. This completes our proof.
\subsection{Coassociativity of the coproduct }\label{app:coalg}
Here we prove that the coproduct defined in equation (\ref{coprod}) is coassociative and satisfies the following condition:
\begin{eqnarray}
\bkt{\Delta\otimes\id}\Delta\sbkt{X}&=&\bkt{\id\otimes\Delta}\Delta\sbkt{X},\ \ \ \forall X\in\mathcal{A}.\label{coass}
\end{eqnarray}
\begin{proof}
We will prove this using induction on the length of the parenthesized words. It is trivial to see that $\Delta$ is coassociative when acting on the words of length $1$. For the induction we assume that it is coassociative acting on words of length $n$. First, we show that it is coassociative on irreducible parenthesized words of length $n+1$ and then we prove the assertion for arbitrary parenthesized words. We use the Sweedler's notation and also drop the summation sign $\sum$ to simplify the notation further. Let $X$ be a parenthesized word of length $n$ then:
\begin{eqnarray}
\Delta\sbkt{X}&=&X_1\otimes X_2,\label{eq:Dx}\\
\bkt{\Delta\otimes\id}\Delta\sbkt{X}&=&\bkt{\id\otimes\Delta}\Delta\sbkt{X}\label{eq:IndAss},
\end{eqnarray}
where equation (\ref{eq:Dx}) is just the simplified Sweedler's notation and equation (\ref{eq:IndAss}) is the induction assumption. Now, consider the parenthesized word $\left(Xx_j\right)$ of length $n+1$. A straightforward computation gives:
\begin{eqnarray}
\bkt{\Delta\otimes\id}\Delta\sbkt{Xx_j}&=&\bkt{\Delta\otimes\id}\bkt{\bkt{Xx_j}\otimes e+\bkt{\id\otimes B_{\bkt{x_j}}}\bkt{X_1\otimes X_2}},\\
&=&\Delta\sbkt{Xx_j}\otimes e + \Delta\sbkt{X_1}\otimes \bkt{X_2x_j},\\
&=&\bkt{Xx_j}\otimes e\otimes e + X_1\otimes \bkt{X_2x_j}\otimes e + \Delta\sbkt{X_1}\otimes \bkt{X_2x_j}.\label{eq:pr1}
\end{eqnarray}
Now, let us compute the RHS of equation (\ref{coass}). By using the definition (\ref{coprodSweed}), we get.
\begin{eqnarray}
\bkt{\id\otimes\Delta}\Delta\sbkt{\bkt{Xx_j}}&=&\bkt{Xx_j}\otimes e\otimes e +X_1\otimes \bkt{X_2x_j}\otimes e + X_1\otimes\sbkt{\bkt{\id\otimes B_{\bkt{x_j}}}\Delta\sbkt{X_2}}.\ \ \ \label{eq:pr2}
\end{eqnarray}
First two terms in the above equation are the same as in equaton (\ref{eq:pr1}). Let's focus on the third term. An important result in this regard is the following.
\begin{eqnarray}
\bkt{\id\otimes\id\otimes B_{\bkt{x_j}}}\bkt{\id\otimes\Delta}\Delta\sbkt{X} &=& \bkt{\id\otimes\id\otimes B_{\bkt{x_j}}}\bkt{\id\otimes\Delta}\bkt{X_1\otimes X_2},\\
&=&\bkt{\id\otimes\id\otimes B_{\bkt{x_j}}}\bkt{X_1\otimes \Delta\sbkt{X_2}},\\
&=& X_1\otimes \sbkt{\bkt{\id\otimes B_{\bkt{x_j}}}\Delta\sbkt{X_2}}.
\end{eqnarray}
Using this we can write:
\begin{eqnarray}
X_1\otimes \sbkt{\bkt{\id\otimes B_{\bkt{x_j}}}\Delta\sbkt{X_2}}&=&\bkt{\id\otimes\id\otimes B_{\bkt{x_j}}}\bkt{\id\otimes\Delta}\Delta\sbkt{X},\\
&=&\bkt{\id\otimes\id\otimes B_{\bkt{x_j}}}\bkt{\Delta\otimes\id}\Delta\sbkt{X}, \\
&=&\bkt{\id\otimes\id\otimes B_{\bkt{x_j}}}\bkt{\Delta\otimes\id}\sbkt{X_1\otimes X_2},\\
&=&\Delta\sbkt{X_1}\otimes \bkt{X_2x_j},
\end{eqnarray}
where, the second equality just follows from the induction assumption (\ref{eq:IndAss}). This is precisely the third term in equation (\ref{eq:pr1}) and this complete the proof of coassociativity for parenthesized words of length $n+1$ an of the form $\bkt{Xx_j}$. Now, for a general parenthesized word $XY$ of length $n+1$, we use the property of the coproduct (\ref{compatcoprod}) to get:
\begin{eqnarray}
\bkt{\id\otimes\Delta}\Delta\sbkt{XY}&=&\bkt{\id\otimes\Delta}\bkt{\Delta\sbkt{X}\Delta\sbkt{Y}},\\
&=&\bkt{\bkt{\id\otimes\Delta}\Delta\sbkt{X}}\bkt{\bkt{\id\otimes\Delta}\Delta\sbkt{Y}},\\
&=&\bkt{\bkt{\Delta\otimes\id}\Delta\sbkt{X}}\bkt{\bkt{\Delta\otimes\id}\Delta\sbkt{Y}},\\
&=&\bkt{\Delta\otimes\id}\bkt{\Delta\sbkt{X}\Delta\sbkt{Y}},\\
&=&\bkt{\Delta\otimes\id}\Delta\sbkt{XY},
\end{eqnarray}
where the first and second lines follow from property (\ref{compatcoprod}), third equality follows fromt the induction assumption. Fourth and fifth lines again follow from (\ref{compatcoprod}). This complete the proof of coassociativity for the coproduct.
\end{proof}
\subsection{Equivalence of definitions of antipode}\label{app:antipodeDef}
In the definition of the antipode, two definitions, (\ref{eq:Sdef1}) and (\ref{eq:Sdef2}), were given. For the antipode to be well defined, these two definitions should be equivalent. We prove this equivalence in the following.
\begin{proof}
We can strip off the parenthesized word $\bkt{Xx_i}$ from the argument of the antipode in equations (\ref{eq:Sdef1}) and (\ref{eq:Sdef2}) and represent antipode as an operator acting on $\mathcal{A}$. Then, we need to show that:
\begin{eqnarray}
-\id-m\sbkt{\bkt{S\otimes\id}P_2 \Delta}\ &=&-\id-\ m\sbkt{\bkt{\id\otimes S}P_2\Delta},\\
-\id-m\sbkt{\bkt{S P_1\otimes P_1} \Delta}\ &=&-\id-\ m\sbkt{\bkt{P_1\otimes SP_1}\Delta}.\label{eq:claim1}
\end{eqnarray}
Both sides still involve the antipode $S$, let us do one more iteration on the both sides. For the left hand side we get:
\begin{eqnarray}
LHS&=& -\id-m\sbkt{\bkt{\bkt{-\id-m\sbkt{\bkt{S P_1\otimes P_1} \Delta}}P_1\otimes P_1} \Delta},\\
&=&-\id+m\sbkt{\bkt{P_1\otimes P_1} \Delta}+m\sbkt{\bkt{\bkt{m\sbkt{\bkt{S P_1\otimes P_1} \Delta}}P_1\otimes P_1} \Delta},\\
&=&-\id+m\sbkt{\bkt{P_1\otimes P_1} \Delta}\nonumber \\ &&+m\sbkt{\bkt{m\otimes\id}\bkt{S\otimes\id\otimes\id}\bkt{P_1\otimes P_1\otimes\id}\bkt{\Delta\otimes\id}\bkt{P_1\otimes P_1}\Delta},\\ 
&=&-\id+m\sbkt{\bkt{P_1\otimes P_1} \Delta}\nonumber \\ &&+m\sbkt{\bkt{m\otimes\id}\bkt{S\otimes\id\otimes\id}\bkt{P_1\otimes P_1\otimes P_1}\bkt{\Delta\otimes\id}\Delta},\label{LHS1}
\end{eqnarray}
where the last equality follows because of the fact that $P_1\otimes P_1 \Delta P_1 = P_1\otimes P_1\Delta$, which is easy to confirm. For the right hand side, a similar computation yields:
\begin{eqnarray}
RHS&=&-\id+m\sbkt{\bkt{P_1\otimes P_1} \Delta}\nonumber \\ &&+m\sbkt{\bkt{\id\otimes m}\bkt{\id\otimes\id\otimes S}\bkt{P_1\otimes P_1\otimes P_1}\bkt{\id\otimes \Delta}\Delta}.\label{RHS1}
\end{eqnarray}
From equations (\ref{LHS1}) and (\ref{RHS1}), we deduce that, to show the equivalence of the two definitions we need to prove the following:
\begin{eqnarray}
&&m\sbkt{\bkt{\id\otimes m}\bkt{\id\otimes\id\otimes S}\bkt{P_1\otimes P_1\otimes P_1}\bkt{\id\otimes \Delta}\Delta}\nonumber \\ &=&
m\sbkt{\bkt{m\otimes\id}\bkt{S\otimes\id\otimes\id}\bkt{P_1\otimes P_1\otimes P_1}\bkt{\Delta\otimes\id}\Delta}
\end{eqnarray}
This is very easy to show using the previously established properties of the coproduct $\Delta$ and the product $m$. Using the coassociativity $\bkt{\Delta\otimes \id}\Delta=\bkt{\id\otimes\Delta}\Delta$, we can freely make the following change in the left side of the above equation:
\begin{eqnarray}
\id\otimes\id\otimes S \to \id\otimes S \otimes \id.
\end{eqnarray}
Similarly, now we make use of the associativity of the product, this implies that $m\bkt{\id\otimes m}=m\bkt{m\otimes \id}$. Using this, we can again move the last two operators in the direct product to the first two places, yielding:
\begin{eqnarray}
\id\otimes S \otimes \id \to S\otimes\id\otimes\id.
\end{eqnarray}
This completes the proof for the equivalence of the two definitions.
\end{proof}
\subsection{Hopf algebra check}\label{app:hopfcheck}
Here, we show that the antipode defined earlier in this article actually satisfies the condition (\ref{hopfCom}).
\begin{proof}
We will do this using induction. For a parenthesized word of length 1, $(x)$, it is easy to see that:
\begin{eqnarray}
E\circ \overline{e}\sbkt{\bkt{x}}=0,
\end{eqnarray} 
and
\begin{eqnarray}
m\sbkt{\bkt{S\otimes \id}\Delta\sbkt{\bkt{x}}}&=&m\sbkt{\bkt{S\otimes\id}\bkt{\bkt{x}\otimes e + e\otimes \bkt{x}}},\\
&=&m\sbkt{-\bkt{x}\otimes e + e\otimes \bkt{x}}=0.
\end{eqnarray}
A similar computation yields
\begin{eqnarray}
m\sbkt{\bkt{\id\otimes S}\Delta\sbkt{\bkt{x}}}=0.
\end{eqnarray}
Let us now assume that the assertion holds for parenthesized words of length $n$, consider an irreducible parenthesized word $\bkt{Xx}$ of length $n+1$. Since the map $P_1$ annihilates $e$, using the Sweedler's notation we can write the antipode of $\bkt{Xx}$ as follows:
\begin{eqnarray}
S\sbkt{\bkt{Xx}}=-\bkt{Xx}-\sum_{X_{1}\neq e}S\sbkt{X_1}\bkt{X_{2}\ x}.
\end{eqnarray}
Now, 
\begin{eqnarray}
\Delta\sbkt{\bkt{Xx}}&=&\bkt{Xx}\otimes e +\sum_{X_1}X_{1}\otimes \bkt{X_2\ x},\\
\bkt{S\otimes \id}\Delta\sbkt{\bkt{Xx}}&=& S\sbkt{\bkt{Xx}}\otimes e + \sum_{X_1}S\sbkt{X_{1}}\otimes \bkt{X_2\ x},\\
&=&-\bkt{Xx}\otimes e-\sum_{X_{1}\neq e}S\sbkt{X_1}\bkt{X_{2}\ x}\otimes e+ \sum_{X_1}S\sbkt{X_{1}}\otimes \bkt{X_2\ x},\\
m\bkt{S\otimes \id}\Delta\sbkt{\bkt{Xx}}&=&-\bkt{Xx}-\sum_{X_{1}\neq e}S\sbkt{X_1}\bkt{X_{2}\ x}+ \sum_{X_1}S\sbkt{X_{1}} \bkt{X_2\ x},\\
&=&-\bkt{Xx}+S\sbkt{e}\bkt{Xx}=0,
\end{eqnarray}
where the last line follows from the fact that then when $X_1=e$, $X_2=X$. 
Now, let us consider the case for $m\sbkt{\bkt{\id\otimes S}\Delta\sbkt{\bkt{Xx}}}$. Due to the equivalence of two definitions (\ref{eq:Sdef1}) and (\ref{eq:Sdef2}), and the properties of $P_2$ we have the following identity:
\begin{eqnarray}
\bkt{S\otimes \id}\sum_{X_1\neq e}X_1\otimes \bkt{X_2 x}= \bkt{\id\otimes S}\sum_{X_1\neq e}X_1\otimes \bkt{X_2 x}.
\end{eqnarray}
Using this, we find that:
\begin{eqnarray}
\bkt{\id\otimes S}\Delta\sbkt{\bkt{Xx}}&=&\bkt{Xx}\otimes e + \bkt{\id\otimes S}\sum_{X_1}X_1\otimes \bkt{X_2 x},
\\
&=& \bkt{Xx}\otimes e + \bkt{S\otimes \id}\sum_{X_1\neq e}X_1\otimes \bkt{X_2 x}+\bkt{\id\otimes S}e\otimes \bkt{Xx},
\\
&=&\bkt{Xx}\otimes e + \sum_{X_1\neq e}S\sbkt{X_1}\otimes \bkt{X_2 x}+\bkt{\id\otimes S}\sbkt{e\otimes \bkt{Xx}},\\
&=& \bkt{Xx}\otimes e + \sum_{X_1\neq e}S\sbkt{X_1}\otimes \bkt{X_2 x}-e\otimes \bkt{Xx}\nonumber \\ &&-\sum_{X_{1}\neq e}e\otimes S\sbkt{X_1}\bkt{X_2x},
\end{eqnarray}
which implies
\begin{eqnarray}
m\sbkt{\bkt{\id\otimes S}\Delta\sbkt{\bkt{Xx}}}&=& 0.
\end{eqnarray}
Since the counit annihilates any parenthesized word we finally conclude that:
\begin{eqnarray}
m\sbkt{\bkt{S\otimes \id}\Delta\sbkt{\bkt{Xx}}}=0=E\circ\overline{e}\sbkt{\bkt{Xx}}.
\end{eqnarray}
For an arbitrary parenthesized word $XY$, due to the induction assumption and the property (\ref{compatcoprod}) of the coproduct, the assertion holds trivially.
This completes our proof. 
\end{proof}

\section{Example}\label{app:example}
Here, we will work out an elementary example which elucidates how all the different elements of the Hopf algebra fit together to give a finite result for a divergent integral. We will use a very simple toy model, defined below:
\begin{eqnarray}
\bkt{x_{j}}\sbkt{c}&\equiv& \int_{c}^{\infty}dy y^{-1-j\epsilon}\equiv I_j,\\
\bkt{Xx_{j}}\sbkt{c}&\equiv& \int_{c}^{\infty}dy y^{-1-j\epsilon}X\sbkt{y},\\
\bkt{x_j}\bkt{x_k}\sbkt{c}&=& I_jI_k,\\
R\sbkt{X\sbkt{c}}&\equiv& X\sbkt{1}.
\end{eqnarray}
It is easy to see that $I_{j}$ is divergent as $\frac{1}{j\epsilon}$. We call the subscript $j$ in $\bkt{x_j}$, the loop order of $\bkt{x_j}$. This toy model is the simplest realization of our Hopf algebra. Let us consider the divergent graph $X=\bkt{\bkt{x_1}\bkt{x_2}x_1}$. Our claim is that the expression $X_r\equiv m\sbkt{\bkt{S_R\otimes \id}\Delta\sbkt{X\sbkt{c}}}$ is a finite integral as expected from our Hopf algebra construction. By making use of the already worked out examples for $\Delta\sbkt{\bkt{\bkt{x_i}\bkt{x_j}x_k}}$, $S_R\sbkt{\bkt{\bkt{x_i}x_j}}$, $S_{R}\sbkt{\bkt{\bkt{x_i}\bkt{x_j}x_k}}$ and the fact $S_{R}\sbkt{XY}=S_R\sbkt{X}S_R\sbkt{Y}$  we find that:
\begin{eqnarray}
X_r&=&X\sbkt{c}-\bkt{x_1}\sbkt{1}\bkt{\bkt{x_2}x_1}\sbkt{c}-\bkt{x_2}\sbkt{1}\bkt{\bkt{x_1}x_1}\sbkt{c}+\bkt{x_1}\sbkt{1}\bkt{x_2}\sbkt{1}\bkt{x_1}\sbkt{c}\nonumber \\ &&-\bkt{\text{first four terms with }c\text{ replaced by }1}.\label{eq:Xr}
\end{eqnarray}
Now, 
\begin{eqnarray}
T_1\equiv X\sbkt{c}&=&\int_{c}^{\infty}dx x^{-1-\epsilon}\int_{x}^{\infty}dy y^{-1-2\epsilon}\int_{x}^{\infty}dz z^{-1-\epsilon},\\
T_2\equiv - \bkt{x_1}\sbkt{1}\bkt{\bkt{x_2}x_1}\sbkt{c}&=&- \int_{c}^{\infty}dx x^{-1-\epsilon}\int_{x}^{\infty}dy y^{-1-2\epsilon}\int_{1}^{\infty}dz z^{-1-\epsilon},\\
T_3\equiv - \bkt{x_2}\sbkt{1}\bkt{\bkt{x_1}x_1}\sbkt{c}&=&-\int_{c}^{\infty}dx x^{-1-\epsilon}\int_{1}^{\infty}dy y^{-1-2\epsilon}\int_{x}^{\infty}dz z^{-1-\epsilon},\\
T_4\equiv \bkt{x_1}\sbkt{1}\bkt{x_2}\sbkt{1}\bkt{x_1}\sbkt{c}&=&\int_{c}^{\infty}dx x^{-1-\epsilon}\int_{1}^{\infty}dy y^{-1-2\epsilon}\int_{1}^{\infty}dz z^{-1-\epsilon}.
\end{eqnarray}
The first two terms can be combined to get:
\begin{eqnarray}
T_1+T_2=-\int_{c}^{\infty}dx x^{-1-\epsilon}\int_{x}^{\infty}dy y^{-1-2\epsilon}\int_{1}^{x}dz z^{-1-\epsilon}.
\end{eqnarray}
The third term can be written as 
\begin{eqnarray}
T_3&=&-\int_{c}^{\infty}dx x^{-1-\epsilon}\int_{1}^{\infty}dy y^{-1-2\epsilon}\bkt{\int_{1}^{\infty}dz z^{-1-\epsilon}-\int_{1}^{x}dz z^{-1-\epsilon}},\\
T_3 &=&-T_4+\int_{c}^{\infty}dx x^{-1-\epsilon}\int_{1}^{\infty}dy y^{-1-2\epsilon}\int_{1}^{x}dz z^{-1-\epsilon}.
\end{eqnarray}
So that the sum of the four terms is:
\begin{eqnarray}
T_1+T_2+T_3+T_4=\int_{c}^{\infty}dx x^{-1-\epsilon}\int_{1}^{x}dy y^{-1-2\epsilon}\int_{1}^{x}dz z^{-1-\epsilon}.
\end{eqnarray}
Plug this in equation (\ref{eq:Xr}) we finally obtain the expression:
\begin{eqnarray}
X_{r}=-\int_{1}^{c}dx x^{-1-\epsilon}\int_{1}^{x}dy y^{-1-2\epsilon}\int_{1}^{x}dz z^{-1-\epsilon},
\end{eqnarray}
which is clearly well defined and finite in the $\epsilon \to 0$ limit. Although this was a very simple example, there should be no hinderance in generalizing this to more realistic QFT examples. If we consider some realistic Feynman graph, our Hopf algebra will renormalize it with the same ease by applying the operator $m\sbkt{\bkt{S\otimes \id}\Delta}$.

\end{document}